\documentclass[aps,showpacs,preprintnumbers,amsmath, amssymb]{revtex4}

\usepackage{float}
\usepackage{graphics,epsfig}
\usepackage{graphicx}
\usepackage{dcolumn}
\usepackage{bm}

\begin{document}
\baselineskip=0.8 cm

\title{{\bf Hair distributions in noncommutative Einstein-Born-Infeld black holes}}
\author{Yan Peng$^{1}$\footnote{yanpengphy@163.com}}
\affiliation{\\$^{1}$ School of Mathematical Sciences, Qufu Normal University, Qufu, Shandong 273165, China}

\vspace*{0.2cm}
\begin{abstract}
\baselineskip=0.6 cm
\begin{center}
{\bf Abstract}
\end{center}

We study hair mass distributions in noncommutative
Einstein-Born-Infeld hairy black holes with non-zero
cosmological constants. We find that
the larger noncommutative parameter makes the hair
easier to condense in the near horizon area.
We also show that Hod's lower bound can be evaded
in the noncommutative gravity.
However, for large black holes
with a non-negative cosmological constant,
Hod's lower hair mass bound almost holds in the sense
that nearly half of the hair lays above the photonsphere.

\end{abstract}

\maketitle
\newpage
\vspace*{0.2cm}

\section{Introduction}

The famous black hole no hair conjecture introduced by
Wheeler \cite{RR,JDB,PW} was motivated by the uniqueness
theorems, establishing that an Einstein-Maxwell black hole can
be completely determined only by the three conserved global charges
associated with Gauss laws, which are the ADM mass M, electric
charge Q and angular momentum J \cite{WI1,WI2,BC,SW,DC1,DC2,JI}.
In accordance with this no hair theorem, stationary black holes indeed
cannot support the existence of scalar fields, massive
vector fields and spinor fields in the exterior
spacetime, for related references see \cite{JE1,JE2,JE3,BM1,BM2,BM3,JH1,JH2,YY1,YY2}.

However, nowadays we are faced with the surprising discovery of various types of
hairy black holes in theories like Einstein-Yang-Mills, Einstein-Skyrme,
Einstein-non-Abelian-Proca, Einstein-Yang-Mills-Higgs,
Einstein-Yang-Mills-Dilaton and kerr black holes with scalar hair,
which cannot be uniquely described by the three conserved charges M, Q and J,
for references please refer to \cite{PBH1}-\cite{H23} and reviews can be found in \cite{H24,H25}.
Recently, a no short hair theorem was proposed as an alternative
to the classical no hair theorem based on the fact that
the hair satisfying the weak energy condition
and the energy-momentum tensor dominant condition must extend above the photonsphere \cite{DNH}.
And it was found that no short scalar hair behaviors also exist
in non-spherically symmetric non-static kerr black holes \cite{SHA}.
It also provided a nice heuristic picture that the formation of hair
is due to the self-interaction which can
bind together the hair below the photonsphere
and hair above the photonsphere relatively distant from the horizon \cite{DNH,SHA}.

Along this line, it is interesting to study the hair distribution
outside the black hole horizon.
For the limit case of the linear Maxwell field,
Hod showed that the region above the photonsphere
contains at least half of the total mass of Maxwell fields
and also found that this lower bound
holds for various genuine hairy black holes
in Einstein-Yang-Mills, Einstein-Skyrme, Einstein-non Abelian-Proca,
Einstein-Yang-Mills-Higgs and Einstein-Yang-Mills-Dilaton systems \cite{SH}.
And Hod further conjectured that the hair mass lower bound exists in all hairy black holes.
In fact, it was found that the non-linear Einstein-Born-Infeld black holes also satisfy
this lower bound that half of the Born-Infeld hair is above the photonsphere \cite{YSM}.
As a further step, we showed that Hod's lower bound holds in asymptotically
dS Einstein-Born-Infeld hairy black holes
and Hod's lower bound can be evaded in the
asymptotically AdS gravity \cite{YP5}.

However, all front calculations were based on commutative spacetimes.
Recently, noncommutative black holes have been studied
on the motivation that noncommutativity is expected to be relevant at the
Planck scale where it is known that usual semiclassical considerations break down.
For example,  modifications to the semiclassical
area law in the noncommutative (NC) spacetime have been obtained \cite{PN1,PN2,PN3,PN4,PN5,PN6}.
Another important motivation to study noncommutative theories is
due to its natural emergence in string theory and some surprising consequences \cite{EWN,NSE,SMN,JGT,DB,KM,YQS}.
In this work, we plan to extend the discussion of hair distributions
to noncommutative spacetimes and also examine whether
Hod's lower hair mass bound holds in noncommutative hairy black holes.

In the following, we introduce noncommutative Einstein-Born-Infeld black holes
and disclose effects of parameters on hair distributions.
We also examine whether Hod's lower bound holds in this noncommutative model.
And we will summarize our main results at the last section.

\section{Hair mass bounds in noncommutative hairy black holes}

In this paper, we choose the background of noncommutative Einstein-Born-Infeld hairy black holes and
the corresponding Lagrangian density with non-zero cosmological constant $\Lambda$ is \cite{LR1,LR2,LR3,LR4}
\begin{eqnarray}\label{AdSBH}
L=R-2\Lambda+4b^2(1-\sqrt{1+\frac{F^{\mu\nu}F_{\mu\nu}}{2b^2}}).
\end{eqnarray}
Here R is the scalar curvature, b is the Born-Infeld factor parameter
and the limit of $b\rightarrow\infty$ corresponds
to the Maxwell field case.

Now we introduce the line element of Einstein-Born-Infeld black holes with
noncommutative mass deformation as follows \cite{PN5,DG1}
\begin{eqnarray}\label{AdSBH}
ds^{2}&=&-f_{EBI}(r)dt^{2}+f(r)_{EBI}^{-1}dr^{2}+r^{2}(d\theta^2+sin^{2}\theta d\phi^{2}).
\end{eqnarray}
The metric function is $f_{EBI}(r)=1-\frac{4M}{\sqrt{\pi}r}\gamma(\frac{3}{2},\frac{r^2}{4\theta})-\Lambda r^2+
\frac{2b^2r^2}{3}(1-\sqrt{1+\frac{Q^2}{b^2r^4}})+\frac{4Q^2}{3r^2}F(\frac{1}{4},\frac{1}{2},\frac{5}{4};-\frac{Q^2}{b^2r^4})$,
where Q is the charge, M is the ADM mass, F is the hypergeometric function satisfying
$(\frac{1}{r}F[\frac{1}{4},\frac{1}{2},\frac{5}{4};-\frac{Q^2}{b^2r^4}])'_{r}=-\frac{1}{\sqrt{r^4+\frac{Q^2}{b^2}}}$
and $\gamma$ is the incomplete gamma function defined as $\gamma(n,z)=\int_{0}^{z}t^{n-1}e^{-t}dt$.
We also label $\theta$ as the parameter
to describe the effect of noncommutativity of spacetime
and the model goes back to the commutative case
in the limit of $\theta\rightarrow 0$.

We should emphasize that the Einstein-Born-Infeld black hole is hairy since the
Born-Infeld factor $b^2$ is associated with no conserved charge \cite{HB1,HB2}.
At infinity, the EBI black hole with $(M,Q,b^2)$ is indistinguishable
from the RN black hole (M,Q), which implies that $b^2$
is considered as a free parameter like color index n for
the EYM hairy black hole \cite{HB3,HB4,HB5}.
Indeed, Breton showed clearly that the BI parameter $b^2$ plays the
same role of color index n for the EYM black hole \cite{HB6}.

The mass in a sphere of radius r is
\begin{eqnarray}\label{AdSBH}
m(r)=M-\frac{\sqrt{\pi}}{\gamma(\frac{3}{2},\frac{r^2}{4\theta})}
[\frac{b^2r^3}{3}(1-\sqrt{1+\frac{Q^2}{b^2r^4}})+\frac{2Q^2}{3r}F(\frac{1}{4},\frac{1}{2},\frac{5}{4};-\frac{Q^2}{b^2r^4})].
\end{eqnarray}

It was found that the black hole horizon $r_{H}$ and
the photonsphere $r_{\delta}$ can be conveniently used to
describe spatial distribution of the matter field outside the horizon \cite{DNH,SH}.
And the spatial distribution of the hair is characterized
by the dimensionless hair mass ratio
$\frac{m^{+}_{hair}}{m^{-}_{hair}}$, where
\begin{eqnarray}\label{AdSBH}
m^{+}_{hair}=M-m(r_{\delta})
\end{eqnarray}
is the hair mass above the photonsphere
and
\begin{eqnarray}\label{AdSBH}
m^{-}_{hair}=m(r_{\delta})-m(r_{H})
\end{eqnarray}
is the hair mass between the horizon and the photonsphere.
The horizon radius $r_{H}$ is determined by $f_{EBI}(r_{H})=0$.
According to the approach in \cite{SH,YY4}, the photon sphere radius $r_{\delta}$
is determined by the relation
\begin{eqnarray}\label{AdSBH}
2f_{EBI}(r_{\delta})-r_{\delta}f'_{EBI}(r_{\delta})=0.
\end{eqnarray}
We cannot analytically obtain the photon sphere radius and the
horizon radius. Here we have to search for the radius with numerical methods.
The hair mass ratio can be expressed as
\begin{eqnarray}\label{AdSBH}
\frac{m^{+}_{hair}}{m^{-}_{hair}}=\frac{M-m(r_{\delta})}{m(r_{\delta})-m(r_{H})}=
\frac{1}{\frac{\gamma(\frac{3}{2},\frac{r_{\delta}^2}{4\theta})
[b^2r_{H}^3(1-\sqrt{1+\frac{Q^2}{b^2r_{H}^3}})
+\frac{2Q^2}{r_{H}}F(\frac{1}{4},\frac{1}{2},\frac{5}{4};-\frac{Q^2}{b^2r_{H}^4})]}
{\gamma(\frac{3}{2},\frac{r_{H}^2}{4\theta})
[b^2r_{\delta}^3(1-\sqrt{1+\frac{Q^2}{b^2r_{\delta}^3}})
+\frac{2Q^2}{r_{\delta}}F(\frac{1}{4},\frac{1}{2},\frac{5}{4};-\frac{Q^2}{b^2r_{\delta}^4})]}-1}.
\end{eqnarray}

It was found that Hod's lower hair mass bound
$\frac{m^{+}_{hair}}{m^{-}_{hair}}\geqslant 1$
holds for various asymptotically flat static hairy black holes
in Einstein-Yang-Mills, Einstein-Skyrme, Einstein-non Abelian-Proca,
Einstein-Yang-Mills-Higgs, Einstein-Yang-Mills-Dilaton
and Einstein-Born-Infeld systems \cite{SH,YSM}.
In fact, this lower bound also exists in asymptotically
dS static Einstein-Born-Infeld hairy black holes \cite{YP5}.
In the following, we extend the discussion to the case of
noncommutative static Einstein-Born-Infeld hairy black holes.

Case I: $\Lambda=0$

We calculate the hair mass ratio in the
noncommutative Einstein-Born-Infeld asymptotically flat black holes.
We firstly show effects of noncommutative parameter on the hair mass ratio.
With $M=1.0$,~$Q=0.8$,~$\Lambda=0$, $b=2$ and various $\theta$,
we obtain smaller ratio with larger $\theta$ as
\begin{eqnarray}\label{AdSBH}
\frac{m^{+}_{EBI}}{m^{-}_{EBI}}\thickapprox 1.779~~~for~~\theta=0.08;
\end{eqnarray}
\begin{eqnarray}\label{AdSBH}
\frac{m^{+}_{EBI}}{m^{-}_{EBI}}\thickapprox 1.740~~~for~~\theta=0.10;
\end{eqnarray}
\begin{eqnarray}\label{AdSBH}
\frac{m^{+}_{EBI}}{m^{-}_{EBI}}\thickapprox 1.588~~~for~~\theta=0.12.
\end{eqnarray}

\begin{figure}[h]
\includegraphics[width=300pt]{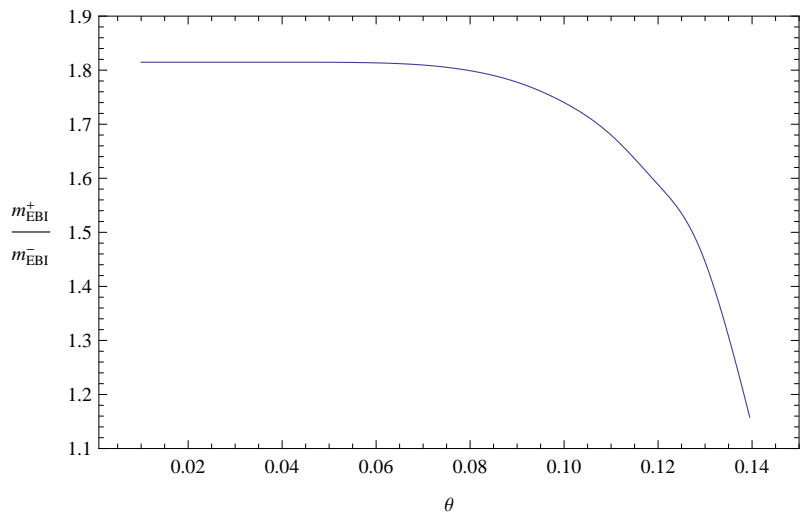}\
\caption{\label{EEntropySoliton} (Color online) We show the hair mass ratio $\frac{m^{+}_{EBI}}{m^{-}_{EBI}}$ as a function of
the noncommutative parameter $\theta$ with $M=1.0$,~$Q=0.8$,~$\Lambda=0$ and $b=2$.}
\end{figure}

\begin{figure}[h]
\includegraphics[width=300pt]{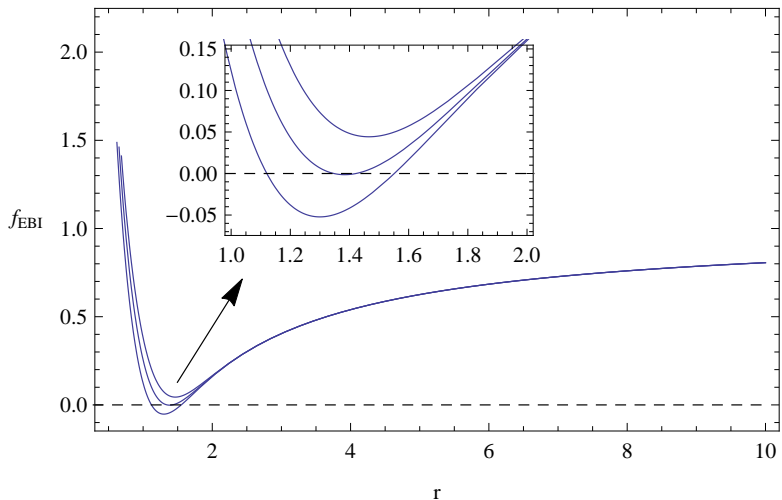}\
\caption{\label{EEntropySoliton} (Color online) We show the metric
function $f_{EBI}$ with $M=1.0$,~$Q=0.8$,~$\Lambda=0$, $b=2$ and various $\theta$.
The three curves from top to bottom correspond to $\theta=0.1600$,~$\theta=0.1395$ and $\theta=0.1200$.}
\end{figure}

We further show $\frac{m^{+}_{EBI}}{m^{-}_{EBI}}$ as a function of $\theta$ in
Fig. 1. It can be easily seen that
$\frac{m^{+}_{EBI}}{m^{-}_{EBI}}$
decreases with respect to $\theta$.
Then, whether very large $\theta$ can evade Hod's lower
bound is a question to be answered.
We numerically show that $\theta$ cannot be arbitrarily large and there exists a
critical noncommutative parameter $\theta_{0}$,
above which the black hole horizon cannot form as can be seen in Fig. 2.
When fixing $M=1.0$,~$Q=0.8$,~$\Lambda=0$ and $b=2$,
the lowest ratio can be reached at $\theta_{0}$ as
\begin{eqnarray}\label{AdSBH}
\frac{m^{+}_{EBI}(\theta_{0})}{m^{-}_{EBI}(\theta_{0})}\thickapprox1.158~~~with~~\theta_{0}\approx0.1395.
\end{eqnarray}
That is to say the noncommunity cannot evade Hod's lower bound in this specific case
of $M=1.0$,~$Q=0.8$,~$\Lambda=0$ and $b=2$.
Now we study the ratio with various Q and $\theta$.
With $M=1.0$,~$\Lambda=0$,~$b=2$ and various $Q$, we can search for $\theta_{0}$
and calculate the minimum ratios at $\theta_{0}$ as
\begin{eqnarray}\label{AdSBH}
\frac{m^{+}_{EBI}(\theta_{0})}{m^{-}_{EBI}(\theta_{0})}\thickapprox1.134~~~for~~\theta_{0}\approx0.1705~~and~~Q=0.7;
\end{eqnarray}
\begin{eqnarray}\label{AdSBH}
\frac{m^{+}_{EBI}(\theta_{0})}{m^{-}_{EBI}(\theta_{0})}\thickapprox1.076~~~for~~\theta_{0}\approx0.1974~~and~~Q=0.6;
\end{eqnarray}
\begin{eqnarray}\label{AdSBH}
\frac{m^{+}_{EBI}(\theta_{0})}{m^{-}_{EBI}(\theta_{0})}\thickapprox0.822~~~for~~\theta_{0}\approx0.2221~~and~~Q=0.5;
\end{eqnarray}
\begin{eqnarray}\label{AdSBH}
\frac{m^{+}_{EBI}(\theta_{0})}{m^{-}_{EBI}(\theta_{0})}\thickapprox0.721~~~for~~\theta_{0}\approx0.2403~~and~~Q=0.4.
\end{eqnarray}
We plot the minimum ratio $\frac{m^{+}_{EBI}(\theta_{0})}{m^{-}_{EBI}(\theta_{0})}$ as a
function of Q with $M=1.0$,~$\Lambda=0$ and $b=2$ in Fig. 3
and $M=1.5$,~$\Lambda=0$ and $b=2$ in Fig. 4.
Here we numerically find that Hod's lower bound can be evaded
in the noncommutative static asymptotically flat Einstein-Born-Infeld black hole model.
We also mention that Hod's lower bound is more likely to
be evaded in the small charge region.
In contrast, it should be emphasized that Hod's lower bound always holds
in the commutative static EBI black holes with non-negative cosmological constants \cite{YSM,YP5}.

\begin{figure}[h]
\includegraphics[width=300pt]{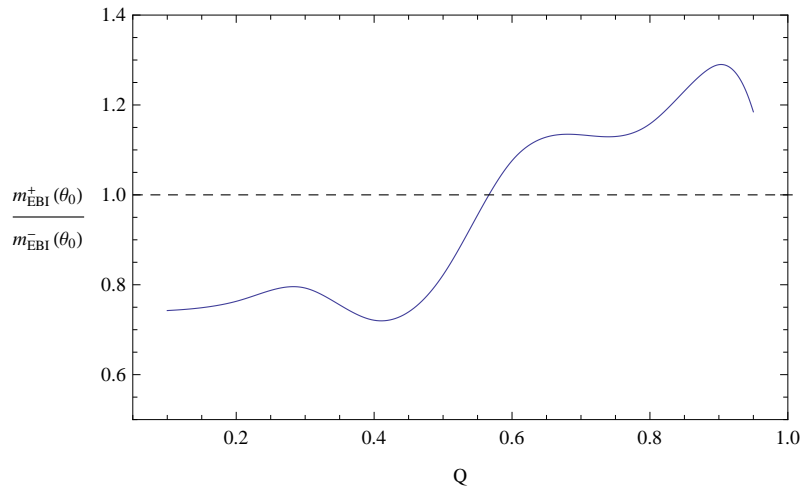}\
\caption{\label{EEntropySoliton} (Color online)We show the minimum hair mass ratio
$\frac{m^{+}_{EBI}(\theta_{0})}{m^{-}_{EBI}(\theta_{0})}$ as a function of
the charge Q with $\Lambda=0$,$b=2$ and $M=1$.}
\end{figure}

\begin{figure}[h]
\includegraphics[width=300pt]{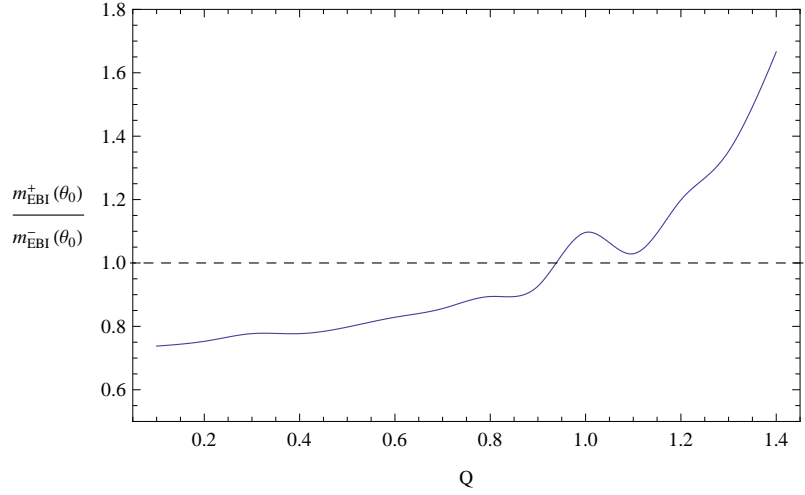}\
\caption{\label{EEntropySoliton} (Color online)We show the minimum hair mass ratio
$\frac{m^{+}_{EBI}(\theta_{0})}{m^{-}_{EBI}(\theta_{0})}$ as a function of
the charge Q with $\Lambda=0$,$b=2$ and $M=1.5$.}
\end{figure}

Case II: $\Lambda>0$

Setting $M=1.0$,~$Q=0.4$,~$b=2$ and $\Lambda=0.0001$
in the asymptotically dS black holes, we get the ratio at $\theta=0.2403$ as
\begin{eqnarray}\label{AdSBH}
\frac{m^{+}_{EBI}}{m^{-}_{EBI}}\thickapprox 0.933< 1.
\end{eqnarray}
In this noncommutative case, we find that Hod's lower bound can be evaded in the
asymptotically dS gravity, which is different from commutative cases
where Hod's lower bound always holds in the asymptotically dS black hole \cite{YP5}.
Setting $M=1.0$,~$Q=0.4$,~$b=2$ and $\Lambda=0.01$, we get
the ratio at $\theta=0.2403$ as
\begin{eqnarray}\label{AdSBH}
\frac{m^{+}_{EBI}}{m^{-}_{EBI}}\thickapprox 1.425\geqslant 1.
\end{eqnarray}
With the relations (15), (16) and (17), we see that large cosmological constants lead to a
mass ratio above Hod's lower mass bound.
With detailed calculation, we find that $\frac{m^{+}_{EBI}}{m^{-}_{EBI}}$
increases with respect to $\Lambda$.

Case III: $\Lambda<0$

Now we show that Hod's bound can also be evaded in the AdS noncommutative background.
With $M=1.0$,~$Q=0.6$,~$b=2$,~$\theta=0.18$ and various $\Lambda$, we find
\begin{eqnarray}\label{AdSBH}
\frac{m^{+}_{EBI}}{m^{-}_{EBI}}\thickapprox 1.448>1~~~for~~~\Lambda=0;
\end{eqnarray}
\begin{eqnarray}\label{AdSBH}
\frac{m^{+}_{EBI}}{m^{-}_{EBI}}\thickapprox 0.726<1~~~for~~~\Lambda=-0.05.
\end{eqnarray}
It means that Hod's bound can be evaded by imposing an AdS boundary.
Due to the confinement of the AdS boundary,
this result is natural and effects of negative cosmological constants on hair
distribution should be qualitatively the same for
other types of black hole hairs.

In the front analysis, we find that Hod's lower bound
can be widely evaded in the noncommutative gravity
and Hod's bound is not such a general property as cases in the commutative case.
However, we will show in the following that Hod's bound almost holds in noncommutative black holes
of large size with non-negative cosmological constants.
Our numerical data shows that the following relation exactly holds with
non-negative cosmological constants and various other parameters
\begin{eqnarray}\label{AdSBH}
1\leqslant\frac{r_{H}^3(1-\sqrt{1+\frac{Q^2}{b^2r_{H}^3}})
+\frac{2Q^2}{r_{H}}F(\frac{1}{4},\frac{1}{2},\frac{5}{4};-\frac{Q^2}{b^2r_{H}^4})}
{r_{\delta}^3(1-\sqrt{1+\frac{Q^2}{b^2r_{\delta}^3}})
+\frac{2Q^2}{r_{\delta}}F(\frac{1}{4},\frac{1}{2},\frac{5}{4};-\frac{Q^2}{b^2r_{\delta}^4})}\leqslant2.
\end{eqnarray}
In the large black hole limit or $r_{\delta}\geqslant r_{H}\gg \theta$, the model
goes back to the commutative case and the relation (20) is
equivalent to Hod's lower bound in the commutative black hole.
Here, we further find that (20) holds beyond the large black hole limit.
According to (20) and the fact that $\frac{\gamma(\frac{3}{2},\frac{r_{\delta}^2}{4\theta})}{\gamma(\frac{3}{2},\frac{r_{H}^2}{4\theta})}\geqslant 1$,
the ratio (7) can be expressed as
\begin{eqnarray}\label{AdSBH}
\frac{m^{+}_{hair}}{m^{-}_{hair}}=
\frac{1}{\frac{\gamma(\frac{3}{2},\frac{r_{\delta}^2}{4\theta})}{\gamma(\frac{3}{2},\frac{r_{H}^2}{4\theta})}\frac{[b^2r_{H}^3(1-\sqrt{1+\frac{Q^2}{b^2r_{H}^3}})
+\frac{2Q^2}{r_{H}}F(\frac{1}{4},\frac{1}{2},\frac{5}{4};-\frac{Q^2}{b^2r_{H}^4})]}
{[b^2r_{\delta}^3(1-\sqrt{1+\frac{Q^2}{b^2r_{\delta}^3}})
+\frac{2Q^2}{r_{\delta}}F(\frac{1}{4},\frac{1}{2},\frac{5}{4};-\frac{Q^2}{b^2r_{\delta}^4})]}-1}\geqslant
\frac{1}{2\cdot\frac{\gamma(\frac{3}{2},\frac{r_{\delta}^2}{4\theta})}{\gamma(\frac{3}{2},\frac{r_{H}^2}{4\theta})}-1}.
\end{eqnarray}

Considering that $\gamma(\frac{3}{2},\frac{r^2}{4\theta})$ increases as a function
of $r$ with values of $\gamma(\frac{3}{2},\frac{r^2}{4\theta})$ in the range $[0,\frac{\sqrt{\pi}}{2}]$,
we obtain lower bound of the ratio in the noncommutative gravity as
\begin{eqnarray}\label{AdSBH}
\frac{m^{+}_{hair}}{m^{-}_{hair}}\geqslant
\frac{1}{2\cdot\frac{\gamma(\frac{3}{2},\frac{r_{\delta}^2}{4\theta})}{\gamma(\frac{3}{2},\frac{r_{H}^2}{4\theta})}-1}
\geqslant \frac{1}{\frac{\sqrt{\pi}}{\gamma(\frac{3}{2},\frac{r_{H}^2}{4\theta})}-1}
=\frac{\gamma(\frac{3}{2},\frac{r_{H}^2}{4\theta})}{\sqrt{\pi}-\gamma(\frac{3}{2},\frac{r_{H}^2}{4\theta})}.
\end{eqnarray}

In the limit of large black hole or $r_{H}\gg\theta$, there is $\frac{m^{+}_{hair}}{m^{-}_{hair}}\geqslant\frac{\gamma(\frac{3}{2},\frac{r_{H}^2}{4\theta})}
{\sqrt{\pi}-\gamma(\frac{3}{2},\frac{r_{H}^2}{4\theta})}\thickapprox 1$.
According to the lower bound of (22), we have
\begin{eqnarray}\label{AdSBH}
\frac{m^{+}_{hair}}{m^{-}_{hair}}\geqslant0.272~~~for~~~\frac{r_{H}^2}{4\theta}\geqslant 1;
\end{eqnarray}
\begin{eqnarray}\label{AdSBH}
\frac{m^{+}_{hair}}{m^{-}_{hair}}\geqslant0.799~~~for~~~\frac{r_{H}^2}{4\theta}\geqslant 3;
\end{eqnarray}
\begin{eqnarray}\label{AdSBH}
\frac{m^{+}_{hair}}{m^{-}_{hair}}\geqslant0.994~~~for~~~\frac{r_{H}^2}{4\theta}\geqslant 7.
\end{eqnarray}
For $\frac{r_{H}^2}{4\theta}\geqslant 7$, nearly half of the hair lays above the photonsphere
and in a certain sense, Hod's bound almost holds for large black holes with non-negative cosmological constants.

According to results in \cite{YP5}, the ratio of (20) is equal to 2 in the limit of large b and small Q.
So the lower bound (22) should be also the approximate formula of the hair mass ratio in the case
of large b and small Q. That is to say $\frac{m^{+}_{hair}}{m^{-}_{hair}}\thickapprox
\frac{\gamma(\frac{3}{2},\frac{r_{H}^2}{4\theta})}{\sqrt{\pi}-\gamma(\frac{3}{2},\frac{r_{H}^2}{4\theta})}$
in the nearly neutral black hole with large b,
which is also well supported by our numerical data. For example, in the case of $M=1.0,~Q=0.1,~b=2$
and $\theta=0.275$, we have $r_{H}\thickapprox1.590$ and
\begin{eqnarray}\label{AdSBH}
\frac{m^{+}_{hair}}{m^{-}_{hair}}\thickapprox0.743;
\end{eqnarray}
\begin{eqnarray}\label{AdSBH}
\frac{\gamma(\frac{3}{2},\frac{r_{H}^2}{4\theta})}
{\sqrt{\pi}-\gamma(\frac{3}{2},\frac{r_{H}^2}{4\theta})}\thickapprox 0.733.
\end{eqnarray}

In summary, we show that Hod's bound can be evaded in the noncommutative
Einstein-Born-Infeld black holes.
We obtain a lower bound (22) expressed with black hole horizon and noncommutative parameters.
And (22) shows that Hod's bound almost holds for large black holes in flat or dS backgrounds.
Since there is also no scalar hair theorem in regular neutral reflecting stars \cite{CP1}
and static scalar fields can condense around charged reflecting stars
\cite{CP2}, it is also very interesting to extend the discussion to the horizonless reflecting star background.
At last, we point out that our discussion of hair distributions are
based on the existence of photon spheres in static black holes.
However, in the rotating case, there is no photon
sphere. There are light rings on the equatorial plane that are different
for co-rotating and counter-rotating orbits.
In this rotating case, light rings are used to determine the effective length of the hair outside
black holes \cite{CP3,CP4}.

\section{Conclusions}

We studied hair distributions of the static spherically symmetric
Einstein-Born-Infeld black hole in the noncommutative geometry.
We used the photonsphere to divide the matter into two parts and
obtained lower bounds of the mass ratio.
We found that the noncommutative parameter makes the hair
easier to condense in the near horizon area.
We further showed that Hod's bound can be evaded in the noncommutative hairy black holes
and Hod's bound is not such a general property as cases in the commutative case.
We also mentioned that Hod's lower bound is more likely to
be evaded in the small charge region.
However, for large black holes
with a non-negative cosmological constant,
Hod's lower hair mass bound almost holds in a sense
that nearly half of the hair lays above the photonsphere.

\begin{acknowledgments}

We would like to thank the anonymous referee for the constructive suggestions to improve the manuscript.
This work was supported by the Shandong Provincial Natural Science Foundation of China under Grant No. ZR2018QA008.

\end{acknowledgments}

\end{document}